\documentclass[12pt]{article}
\usepackage{amssymb, amscd, amsmath, amsthm, epsf, psfig, latexsym}
\usepackage{amsfonts}
\usepackage[dvips]{graphicx}

\newtheorem{thm}{Theorem}[section]

\newtheorem{lemm}[thm]{Lemma}
\newtheorem{SC}[thm]{Scholium}

\newtheorem{prop}[thm]{Proposition}

\newtheorem{definition}{Definition}

\newcommand{\C}{{\mathbb C}}
\newcommand{\Q}{{\Bbb Q}}

\newcommand{\RI}{{\rightharpoonup}}

\def\tn{\textnormal}

\def\U{{\bf U}}

\begin{document}

\title{Simulation of topological field theories by quantum computers}

\author{Michael H. Freedman\footnote{Microsoft
Research, One Microsoft Way, Redmond, WA 98052-6399
},  Alexei Kitaev\footnote{Microsoft
Research, One Microsoft Way, Redmond, WA 98052-6399
(On leave from Landau Institute for Theoretical Physics (Moscow)},
and Zhenghan Wang\footnote{Indiana University, Dept. of Math., Bloomington,
IN 47405}}

\maketitle

\begin{abstract}
Quantum computers will work by evolving a high tensor power of a
small (e.g. two) dimensional Hilbert space by local gates, which
can be implemented by applying a local Hamiltonian $H$ for a time
${t}$. In contrast to this quantum engineering, the most abstract
reaches of theoretical physics has spawned $\lq\lq$topological
models" having a finite dimensional internal state space with no
natural tensor product structure and in which the evolution of the
state is discrete, $H \equiv 0$. These are called topological
quantum filed theories (TQFTs). These exotic physical systems are
proved to be efficiently simulated on a quantum computer.  The
conclusion is two-fold:
\begin{enumerate}
\item TQFTs cannot be used to define a model of computation stronger
than the usual quantum model $\lq\lq BQP$."
\item TQFTs provide a
radically different way of looking at quantum computation.  The
rich mathematical structure of TQFTs might suggest a new quantum
algorithm.
\end{enumerate}
\end{abstract}

\newpage

\section{Introduction}

A \underline{topological quantum field theory} (TQFT) is a
mathematical abstraction, which codifies topological themes in
conformal field theory and Chern-Simons theory. The strictly
$2-$dimensional part of a TQFT is called a \underline{topological}
\underline{modular functor} (TMF).  It (essentially) assigns a
finite dimensional complex Hilbert space $V(\Sigma)$ to each
surface $\Sigma$ and to any (self)-diffeomorphism $h$ of a surface
a linear (auto)morphism $V(h): V(\Sigma) \rightarrow V (\Sigma')$.
We restrict attention to unitary topological modular functors
(UTMF) and show that a quantum computer can efficiently simulate
transformations of any UTMF as a transformation on its
computational state space. We should emphasize that both sides of
our discussion are at present theoretical:  the quantum computer
which performs our simulation is also a mathematical abstraction -
the \underline{quantum circuit model} (QCM) [D][Y].  Very serious
proposals exist for realizing this model, perhaps in silicon, e.g.
[Ka], but we will not treat this aspect.

There is a marked analogy between  the development of the QCM from
1982 Feynman [Fey] to the present, and the development of
recursive function theory in 1930's and 1940's.  At the close of
the earlier period, $\lq\lq$Church's thesis"  proclaimed the
uniqueness of all models of (classical) calculation: recursive
function theory, Turing machine, $\lambda$-calculus, etc.... The
present paper can be viewed as supporting a similar status for QCM
as $\it{the}$ inherently quantum mechanical model of calculation.
The modern reconsideration of computation is founded on the
distinction between \underline{polynomial time} and
\underline{slower} algorithms.  Of course, all functions computed
in the QCM can be computed classically, but probably not in
comparable time. Assigning to an integer its factors, while
polynomial time in QCM [Sh] is nearly exponential time $\sim
e^{n^{1/3}}$ according to the most refined classical algorithms.
The origin of this paper is in thought [Fr] that since ordinary
quantum mechanics appears to confer a substantial speed up over
classical calculations, that some principle borrowed from the
early, string, universe might go still further.  Each TQFT is an
instance of this question since their discrete topological nature
lends itself to translation into computer science. We answer here
in the negative by showing that for a unitary TQFT, the
transformations $V(h)$ have a hidden poly-local structure.
Mathematically, $V(h)$ can be realized as the restriction to an
invariant subspace of a transformation $\prod g _{i}$ on the state
space of a quantum computer where each $g_{i}$ is a
\underline{gate} and the length of the composition is linear in
the length of $h$ as a word in the standard generators,
$\lq\lq$Dehn twists" of the \underline{mapping class group} =
diffeomorphisms $(\Sigma)$/identity component.  Thus, we add
evidence to the unicity of the QCM. Several variants and
antecedents of QCM, including quantum Turing machines, have
previously be shown equivalent (with and without environmental
errors)[Y].

From a physical standpoint, the QCM derives from Schrodinger's
equation as described by Feynman [Fey] and Lloyd [Ll].  Let us
introduce the model. Given a decision problem, the first or
\underline{classical phase} of the QCM is a classical program,
which designs a \underline{quantum circuit} to $\lq\lq$solve"
\underline{instances} of the decision problem of length $n$. A
quantum circuit is a composition $\U_n$ of operators or
\underline{gates} $g_{i} \in  \U(2)$ or $ \U(4)$ taken from some
fixed list of rapidly computable
 matrices\footnote{the $i$-th digit of each entry should be
computable in $poly(i)$ time}, e.g. having algebraic entries.  The
following short list suffices to efficiently approximate any other
choice of gates [Ki]:

\begin{gather*}
\left\{
\begin{vmatrix} 0 & 1 \\ 1 & 0\end{vmatrix}, \quad
\begin{vmatrix} 1 & 0 \\ 0 & i\end{vmatrix}, \quad \text{and} \,\,
\begin{vmatrix} 1 & 0 & 0 & 0\\ 0 & 1 & 0 & 0 \\ 0 & 0 & \frac{1}{\sqrt{2}}
&
 \frac{1}{\sqrt{2}} \\
0 & 0 & \frac{1}{\sqrt{2}} &\frac{-1}{\sqrt{2}}\end{vmatrix}
\right\}.
\end{gather*}

The \underline{gates} are applied on some tensor power space
$(\C^2)^{\otimes k (n)}$ of $\lq\lq$k qubits" and models a local
transformation on a system of $k$ spin $\frac{1}{2}$ particles.
The gate $g$ acts as the identity on all but one or two tensor
factors where it acts as a matrix as above. This is the middle or
\underline{quantum phase} of the algorithm. The final phase is to
perform a local von Neumann measurement on a final state
$\psi_{\text{final}} = \U_{n} (\psi_{\text{initial}})$ (or a
commuting family of the same) to extract a probabilistic answer to
the decision problem.  (The initial states'
$\psi_{\text{initial}}$ must also be locally constructed).  In
this phase, we could declare that observing a certain eigenvalue
with probability $\geq \frac{2}{3}$ means $\lq\lq$yes."  We are
interested only in the case where the classical phase of circuit
design and the length of the designed circuit are both smaller
than some polynomial in $n$. Decision problems which can be solved
in this way are said to be in the computational class BQP:
\underline{bounded-error quantum polynomial}.  The use of $\C^2$,
the $\lq\lq$qubit", is merely a convenience, any decomposition
into factors of bounded dimension gives an equivalent theory. We
say $\U$ is a quantum circuit over $\C^p$ if all tensor factors
have dimension = $p$.

Following Lloyd [Ll], note that if a finite dimensional quantum
system, say $(\C^{2})^{\otimes k}$, evolves by a Hamiltonian $H$,
it is physically reasonable to assert that $H$ is
\underline{poly-local}, $H =\overset{L}{\underset{i=1}{\sum}}
H_\ell,$ where the sum has $\leq$ $poly(k)$ terms and each $H_\ell
= \overset{\sim}{H}_{\ell} \otimes \text{id}$, where
$\overset{\sim}{H}_{\ell}$ acts nontrivially only on a bounded
number (often just two) qubits and as the identity on the
remaining tensor factors.  Now setting Plank's constant $h = 1$,
the time evolution is given by Schrodinger's equation: $\U_t =
e^{2 \pi i t H}$ whereas gates can rapidly approximate [Ki] any
local transformation of the form $ e^{2 \pi i t H_\ell}.$ Only the
nonabelian nature of the unitary group prevents us from
approximating $\U_t$ directly from as
$\overset{L}{\underset{i=1}{\Pi}} e^{2 \pi i H_\ell}$. However, by
the Trotter formula:
\[ \bigl(e^{A/n + B/n}\bigr)^n = e^{A+B} + \mathcal{O}
\biggl(\frac{1}{n}\biggr), \] where the error $\Omega$ is measured
in the operator norm.  Thus, there is a good approximation to
$\U_t$ as a product of gates:
\[\U_t = \bigl(e^{2\pi i
\frac{t}{n} H_{1}} \dots e^{2\pi i \frac{t}{n} H_L}\bigr)^n +
L^{2} \cdot \mathcal{O} \biggl(\frac{1}{n}\biggr).\]

Because of the rapid approximation result of [Ki], in what
follows, we will not discuss quantum circuits restricted to any
small generating set as in the example above, rather we will
permit a $2 \times 2$ or $4 \times 4$ unitary matrix with
algebraic number entries to appear as a gate.

In contrast to the systems considered by Lloyd, the Hamiltonian in
a topological theory vanishes identically, $H=0$, a different
argument - the substance of this paper - is needed to construct a
simulation. The reader may wonder how a theory  with vanishing $H$
can exhibit nontrivial unitary transformations.  The answer lies
in the Feynman path-integral approach to QFT.  When the theory is
constructed from a Lagrangian (functional on the classical fields
of the theory), which only involves first derivatives in time, the
Legendre transform is identically zero [At], but may nevertheless
have nontrivial global features as in the Aharonov-Bohm effect.

Before defining the mathematical notions, we would make two
comments.  First, the converse to the theorem is an open question:
Can some UTMF efficiently simulate a universal quantum computer?
Fault tolerantly?  We would conjecture the answer is yes to both
questions. Second, we would like to suggest that the theorem may
be viewed as a positive result for computation.  Modular functors,
because of their rich mathematical structure, may serve as higher
order language for constructing a new quantum algorithm.  In [Fr],
it is observed that the transformations of UTMF's can readily
produce state vectors whose coordinates are computationally
difficult evaluations of the Jones and Tutte polynomials.  The
same is now known for the state vector of a quantum computer, but
the question of whether any useful part of this information can be
made to survive the measurement phase of quantum computation is
open.

We would like to thank Greg Kupperberg and Kevin Walker for many
stimulating discussions on the material presented here.

\section{Simulating Modular Functors}

We adopt the axiomatization of [Wa] or [T] to which we refer for
details.  Also see, Atiyah [At], Segal [Se], and Witten [Wi].

A \underline{surface} is a compact oriented $2-$manifold with
parameterized boundaries.  Each boundary component has a label
from a finite set \break $\mathcal{L}= \{1, a, b, c, \dots\}$ with
involution $\widehat{\,\,}$, $1 = \widehat{1}$. In examples,
labels might be representations of a quantum group up to a given
level or positive energy representations of a loop group, or some
other algebraic construct. Technically, to avoid projective
ambiguities each surface $\Sigma$ is provided with a Lagrangian
subspace $L \subset H_{1} (\Sigma; Q)$ and each diffeomorphism
$f:\Sigma \rightarrow \Sigma'$ is provided with an integer
$\lq\lq$framing/signature" so the dynamics of the theory is
actually given by a central extension of the mapping class group.
Since these extended structures are irrelevant to our development,
we suppress them from the notation. We use the letter $\ell$ below
to indicate a label set for all boundary components, or in some
cases, those boundary components without a specified letter as
label.

\begin{definition}
\underline{A unitary topological modular functor} (UTMF) is a
functor $V$ from the category of (labeled surfaces with fixed
boundary parameterizations, label preserving diffeomorphisms which
commute with boundary parameterizations) to (finite dimensional
complex Hilbert spaces, unitary transformations) which satisfies:

\begin{enumerate}
\item Disjoint union axiom:  $V(Y_1 \amalg Y_2, \ell_1 \amalg  \ell_2)=
V(Y_1, \ell_1) \otimes V(Y_2, \ell_2)$.
\item Gluing axiom:  let $Y_g$ arise from $Y$ by gluing together a pair of
boundary circles with dual labels, ${x}$ glues to ${\widehat{x}}$,
then $$V(Y_g, \ell)= \bigoplus_{x \epsilon \mathcal{L}} V (Y,
(\ell, {x}, {\widehat{x}}).$$

\item Duality axiom: reversing the orientation of $Y$ and applying
 $\widehat{\,\,}$
to labels corresponds to replacing $V$ by $V^\ast$. Evaluation
must obey certain naturality conditions with respect to gluing and
the action of the various mapping class groups.
\item Empty surface axiom:  $V(\phi) \cong \C$
\item Disk axiom: $V_a = V(D, a) \cong \begin{cases} \C,& \text{if}\quad a =
1
\\ 0,& \text{if}\quad a \neq 1 \end{cases}$
\item Annulus axiom:  $V_{a, b} = V \bigl( A, (a, b) \bigr) \cong
\begin{cases} \C,
& \text{if} \quad a=\widehat{b} \\ 0, & \text{if} \quad a \neq
\widehat{b} \end{cases}$
\item Algebraic axiom:  The basic data, the mapping class group actions and
the maps $F$ and $S$ explained in the proof (from which $V$ may be
reconstructed if the Moore and Seiberg conditions are satisfied,
see [MS] or [Wa] 6.4,  1-14) is algebraic over $\Q$ for some bases
in of $V_a$, $V_{a, \widehat{a}}$,  and $V_{abc}$, where $V_{abc}$
denote
   $V\bigl(P, (a, b,c)\bigr)$ for a (compact)
   $3$-punctured sphere $P$.
$3$-punctured spheres are also called {\it pants}.
\end{enumerate}
\end{definition}

\underline{\bf{Comments}}:
\renewcommand{\labelenumi}{\theenumi)}
\begin{enumerate}
\item
From the gluing axiom, $V$ may be extended via dissection from
simple pieces $D$, $A$, and $P$ to general surfaces $\Sigma$. But
$V(\Sigma)$ must be canonically defined: this looks quite
difficult to arrange and it is remarkable that any nontrivial
examples of UTMFs exist. \label{w1}
\item
The algebraic axiom is usually omitted, but holds for all known
examples.  We include it to avoid trivialities such as a UTMF
where action by, say, a boundary twist is multiplication by a real
number whose binary expansion encodes a difficult or even
uncomputable function:  e.g. the $i^{\text{th}}$ bit is 0 iff the
$i^{\text{th}}$ Turing machine halts.  If there are nontrivial
parameter families of UTMF's, such nonsensical examples must arise
- although they could not be algebraically specified. In the
context of bounded accuracy for the operation of diffeomorphisms
$V(h)$, axiom 7 may be dropped (and simulation by bounded accuracy
quantum circuits still obtained), but we prefer to work in the
exact context since in a purely topological theory exactness is
not implausible.
\item
Axiom 2 will be particularly important in the context of a
\underline{pants} \break \underline{decomposition} of a surface
$\Sigma$. This is a division of $\Sigma$ into a collection of
compact surfaces $P$ having the topology of $3-$punctured spheres
and meeting only in their boundary components which we call
$\lq\lq$cuffs."
\end{enumerate}

\begin{definition}A quantum circuit $\U: (\C^{p})^{\otimes k }
\rightarrow (\C^{p})^{\otimes k} =: W$ is said to simulate on $W$
(exactly) a unitary transformation $\tau: S \rightarrow S$ if
there is a $\C$-linear imbedding $i:S\subset(\C^{p})^{\otimes k}$
invariant under $\U$ so that $\U \circ i= i\circ \tau$. The
imbedding is said to intertwine $\tau$ and $\U$. We also require
that $i$ be computable on a basis in poly(k) time.
\end{definition}

Since we prove efficient simulation of the topological dynamics
for UTMFs $V$, it is redundant to dwell on $\lq\lq$measurement"
within V, but to complete the computational model, we can posit
von Neumann type measurement with respect to any efficiently
computable frame $\mathcal{F}$ in $V_{abc}$.  The space $\C^p$
above, later denoted $X=\C^p$, is defined by $X := \underset{(a,
b, c)\in \mathcal{L}^3}{\oplus} V_{abc}$ and the computational
space $W:= X^{\otimes k}$.  We have set $S:=V(\Sigma)$ and assumed
$\Sigma$ is divided into $k \,\, \, \lq\lq$pants," i.e. Euler
class $(\Sigma) = -k$. Any frame $\mathcal{F}$ extends to a frame
for $V(\Sigma)$ via the gluing axiom once a pants decomposition of
$\Sigma$ is specified. Thus, measurement in $V$ becomes a
restriction of measurement in $W$. It may be physically more
natural to restrict the allowable measurements on $V(\Sigma)$ to
cutting along a simple closed curve $\gamma$ and measuring the
label which appears.  Mathematically, this amounts to transforming
to a pants decomposition with $\gamma$ as one of its decomposition
or $\lq\lq$cuff" curves and then positing a Hermitian operator
with eigenspaces equal to the summands of $V(\Sigma)$
corresponding under the gluing axiom to labels $x$ on $\gamma^{-}$
and $\widehat{x}$ on $\gamma^{+}$, $x \epsilon \mathcal{L}$.

A labeled surface $(\Sigma, \ell)$ determines a mapping class
group $\mathcal{M}=\mathcal{M}(\Sigma, \ell)=\lq\lq$isotopy
classes of orientation preserving diffeomorphisms of $\Sigma$
preserving labels and commuting with boundary parameterization."
For example, in the case of an $n$-punctured sphere with all
labels equal (distinct), $\mathcal{M}={\rm SFB}(n)$, the spherical
framed braid group $\big(\mathcal{M}={\rm PSFB}(n)$, the pure
spherical framed braid group$\big)$. To prove the theorem below,
we will need to describe a generating set $\mathcal{S}$ for the
various $\mathcal{M}$'s and within $\mathcal{S}$ chains of
\underline{elementary moves} which will allow us to prepare to
apply any $s_2 \in \mathcal{S}$ subsequent to having applied $s_1
\in \mathcal{S}$.

Each $\mathcal{M}$ is generated by \underline{Dehn-twists} and
\underline{braid-moves} (See [B]).  A Dehn-twist $D_\gamma$ is
specified by drawing a simple closed curve (s.c.c.) $\gamma$ on
$\Sigma$, cutting along $\gamma$, twisting $2 \pi$ to the right
along $\gamma$ and then regluing. A braid-move $B_\delta$ will
occur only when a s.c.c. $\delta$ cobounds a pair of pants with
two boundary components of $\Sigma$: If the labels of the boundary
components are equal then $B_\delta$ braids them by a right
$\pi$-twist. In the case that all labels equal, there is a rather
short list of $D$ and $B$ generators indicated in Figure 1 below.
Also sketched in Figure 1 is a pants decomposition of diameter
$=\bigl(\mathcal{O}\log b_{1} (\Sigma)\bigr)$, meaning the graph
dual to the pants decomposition has diameter order $\log$ the
first Betti number of $\Sigma$.

\vskip.2in \epsfxsize=5.5in \centerline{\epsfbox{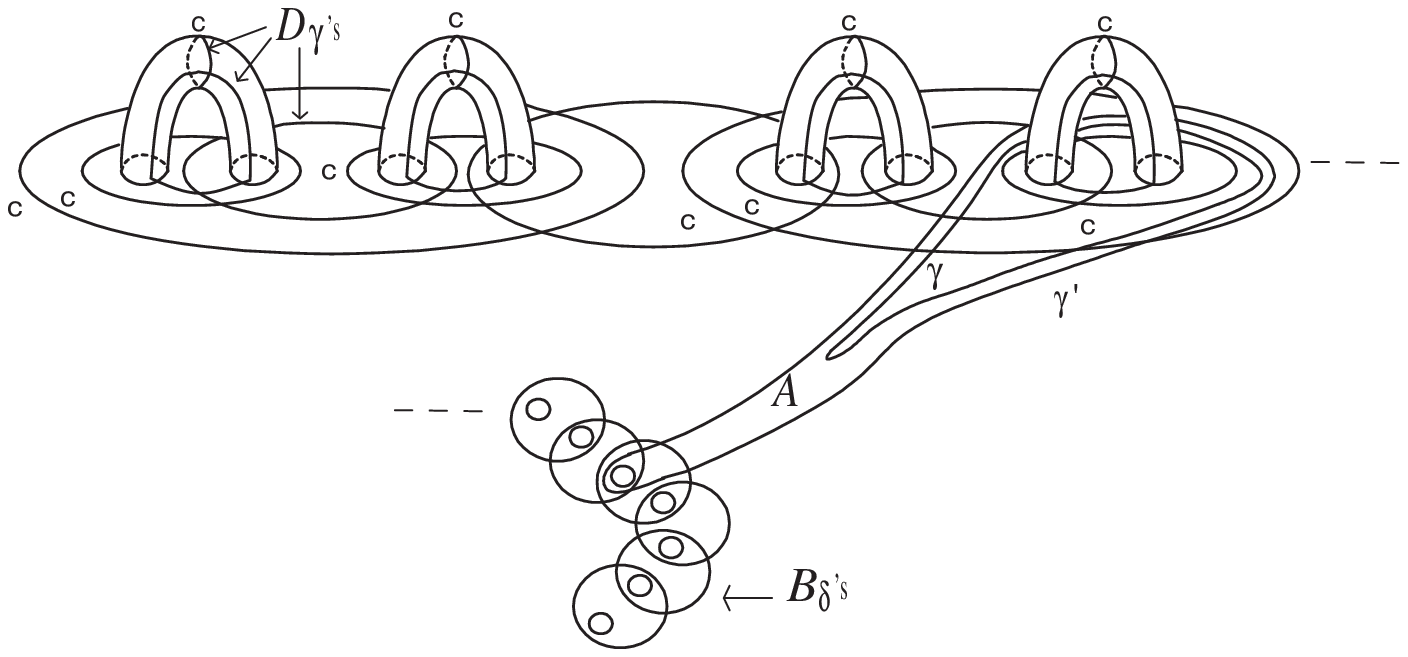}}
\centerline{Figure 1.} \vskip.2in

The s.c.c. $\gamma(\;\delta)$ label Dehn (braid) generators
$D_\gamma\;\; (\text{and} \;\; B_\delta)$. Figure 1 contains a
punctured annulus $A$;  note that the composition of oppositely
oriented Dehn twists along the two $\lq\lq$long" components of
$\partial A$, $\gamma$ and $\gamma'$ yield a diffeomorphism which
moves the punctures about the loop $\gamma$.  The figure
implicitly contains such an $A$ for each $(\gamma, p)$, where $p$
is a \underline{preferred puncture}. The $\gamma$ curves come in
three types:
\renewcommand{\labelenumi}{(\theenumi)}
\begin{enumerate}
\item The loops at the top of the handles which are curves
($\lq\lq$cuffs") of the pants decomposition,
\item loops dual to type 1, and
\item loops running under adjacent pairs of handles (which cut
through up to $\mathcal{O}\bigl(\log(b_{1} \Sigma)\bigr)$ many
cuffs). (See Figure 1, where cuffs are marked by a $\lq\lq$c".)
\end{enumerate}

Each punctured annulus $A$ is determined as a neighborhood (of a
s.c.c. $\gamma$ union an arc $\eta$ from $\gamma$ to $p$).  To
achieve general motions of $p$ around $\Sigma$, we require these
arcs to be $\lq\lq$standard" so that for each $p$, $\pi_1
(\Sigma^{\widehat{}},p)$ is generated by  $\{ \eta \cdot \gamma
\cdot {\eta}^{-1}\}$, where $\Sigma^{\widehat{}} = \Sigma$ with
punctures filled by disks, and the disk corresponding to $p$
serving as a base point. This list of generators is only linear in
the first Betti number of $\Sigma$.

In the presence of distinct labels,  many of the $B_\delta$ are
illegal (they permute unequal labels).  In this case,
quadratically many generators are required.  Figure 2 displays the
replacements for the $B$'s, and additional $A$'s and $D$'s .

\vskip.2in \epsfxsize=3.5in \centerline{\epsfbox{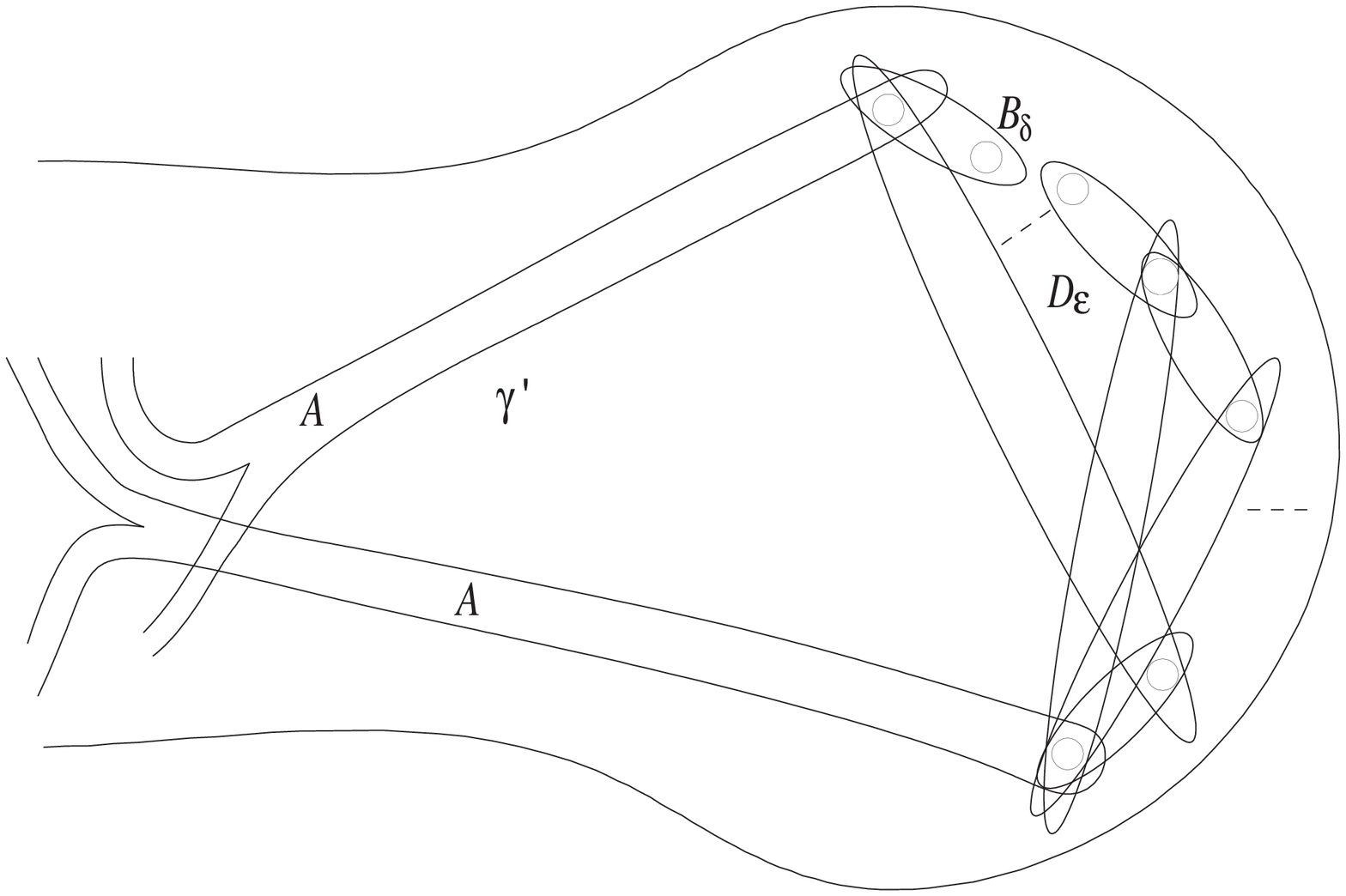}}
\centerline{Figure 2.} \vskip.2in

Figure 2 shows a collection of $B$'s sufficient to effect
arbitrary braiding \underline{within} each commonly-labeled subset
of punctures, a quadratically large collections of new Dehn curves
$\{\epsilon\}$ allowing a full twist between any pair of
distinctly labeled punctures, (If the punctures are arranged along
a convex arc of the Euclidean cell in $\Sigma$, then each
$\epsilon$ will be the boundary of a narrow neighborhood of the
straight line segment joining pairs of dissimilarly labeled
punctures.) and finally a collection of punctured annuli, which
enable one puncture $p_i$ from each label - constant subset to be
carried around each free homotopy class from
$\{\gamma\}$(respecting the previous generation condition for
$\pi_1 (\Sigma^{\widehat{}}, p_i)$.

Thus for distinct labels the generating sets are built from curves
of type $\gamma, \gamma^{\prime}, \epsilon$ and $\delta$ by Dehn
twists around $\gamma,\gamma^{\prime}$, and $\epsilon$,  braid
moves around $\delta$. Denote by $\omega$, any such curve:
$\omega\in \Omega= \bigl\{ \{\gamma\} \cup\{\gamma^{\prime}\} \cup
\{\epsilon\}\cup \{\delta\} \bigr\}$.

Since various $\omega'$s intersect, it is not possible to realize
all $\omega$ simultaneously as cuffs in a pants decomposition.
However, we can start with the $\lq\lq$base point" pants
decomposition $\mathcal{D}$ indicated in Figure 1 (note $\gamma$
of type(1) are cuffs in $\mathcal{D}$, but $\gamma$ of types (2)
or (3) are not) and for any $\omega$ find a short path of
elementary moves: $F$ and $S$ (defined below) to a pants
decomposition $\mathcal{D}_\omega$ containing $\omega$ as a cuff.

\begin{lemm}Assume $\Sigma \neq S^{2}$, disk, or annulus, and
$\mathcal{D}$ the standard pants decomposition sketched in Figure
1.  Any $\omega$ as above, can be deformed through
$\mathcal{O}\bigl(\log b_{1} (\Sigma)\bigr)\,\,F$ and $S$ moves to
a pants decomposition $\mathcal{D}_{\omega}$ in which $\omega$ is
a cuff.
\end{lemm}

We postpone the proof of the lemma and the definition of its terms
until we are partly into the proof of the theorem and have some
experience passing between pants decompositions.

\begin{thm}Suppose $V$ is a UTMF and $h:\Sigma\rightarrow\Sigma$ is a
diffeomorphism of length $n$ in the standard generators for the
mapping class group of $\Sigma$ described above (See Figures 1 and
2). Then there are constants depending only on $V$, $c=c(V)$ and
$p=p(V)$ such that $V(h): V(\Sigma)\rightarrow V(\Sigma)$ is
simulated (exactly) by a quantum circuit operating on
$\lq\lq$qupits" $\C^p$ of length $\leq c \cdot n\cdot \log
b_1(\Sigma)$.
\end{thm}

The collection $\{\text{cuffs}\}$ refers to the circles along
which the pants decomposition decomposes; the $\lq\lq$seams" are
additional arcs, three per pant which cut the pant into two
hexagons.  Technically, we will need each pant in $\mathcal{D}$ to
be parameterized by a fixed $3$-punctured sphere so these seams
are part of the data in $\mathcal{D}$; For simplicity, we choose
seams to minimize the number of intersections with $\{\omega\}$.

The theorem may be extended to cover a more general form of input.
The original algorithm [L] which writes a $\mathcal{D}_\alpha$,
$\alpha$ a s.c.c., as a word in standard generators
$\mathcal{D}_\gamma$ is super-exponential.  We define the
combinatorial length of $\alpha$, $\ell(\alpha)$, to be the
minimum number of intersections as we vary $\alpha$ by isotopy of
$\alpha$ with $\{\text{cuffs}\} \cup \{\text{seams}\}$. The best
upper-bound (known to the authors) to the length $L$ of
$\mathcal{D}_\alpha$ as a word in the mapping class group spanned
by a fixed generating set is of the form $L(\mathcal{D}_\alpha)<$
super-exponential function $f(\ell)$.  For this reason, we
consider as input $V(h)$, where $h$ is a composition of $k$ Dehn
twists on $\alpha_{1}, \dots \alpha_{k}$ and $j$ braid moves along
$\beta_{1}, \dots \beta_{j}$ in any order. Then $V(h)$ is costed
as the sum of the combinatorial length of the simple closed curves
needed to write $h$ as Dehn twists and braid moves within the
mapping class group, $\ell(h)
:=\overset{j}{\underset{i=1}{\Sigma}} \ell(\beta_{i}) +
\overset{k}{\underset{i=1}{\Sigma}} \ell(\alpha_{i})$.  We obtain
the following extension of the theorem.
\\

\noindent{\bf Extension:} \footnote{Lee Mosher has informed us
that the existence of linear bound $f(\ell)$ (but without control
of the constants) follows at least for closed and single punctured
surfaces from his two papers [M1] and [M2].} The map $h_\ast:
V(\Sigma)\longrightarrow V(\Sigma)$ is exactly simulated by a
quantum circuit QC with length (QC) $\leq 11 \ell(h)$ composed of
algebraic $1$ and $2-$qupit $\C^p$ gates.
\\
\\
\noindent{\bf Pre-Proof:} Some physical comments will motivate the proof.
$V(\Sigma)$ are quantized gauge fields on $\Sigma$ (with a
boundary condition given by labels $\ell$) and can be regarded as
a finite dimensional space of internal symmetries.  This is most
clear when genus $(\Sigma)= 0\, ,\Sigma$ is a punctured sphere,
the labeled punctures are $\lq\lq$anyons" [Wil] and the relevant
mapping class group is the \underline{braid group} which moves the
punctures around the surface of the sphere. An internal state
$\psi \epsilon V(\Sigma)$ is transformed to $\U(b) \psi \in
V(\Sigma)$ under the functorial representation of the braid group.
For $\U(b)$ to be defined the braiding must be $\lq\lq$complete"
in the sense that the punctures (anyons) must return setwise to
their initial position.  Infinitesimally, the braiding defines a
Hamiltonian $\overline{H}$ on $V(\Sigma) \otimes E$ where $E$ is
an infinite dimensional Hilbert space which encodes the position
of the anyons.  The projection of $\overline{H}$ into $V(\Sigma)$
vanishes which is consistent with the general covariance of
topological theories.  Nevertheless, when the braid is complete,
the evolution $\overline{\U}$ of $\overline{H}$ will leave
$V(\Sigma)$ invariant and it is $\overline{\U}|_{V(\Sigma)} = \U$
which we will simulate.  Anyons inherently reflect nonlocal
entanglement so it is not to be expected that $V(\Sigma)$ has any
(natural) tensor decomposition and none are observed in
interesting examples.  Thus, simulation of $\U$ as an invariant
subspace of a tensor product $(\C^p)^{\otimes k}$ is the best
result we can expect.  The mathematical proof will loosely follow
the physical intuition of evolution in a super-space by defining,
in the braid case (identical labels and genus $=0$), two distinct
imbeddings $\lq\lq$odd" and $\lq\lq$even," $ V(\Sigma)
\genfrac{}{}{0pt}{}{\stackrel{\mathrm{odd}}{\longrightarrow}}
{\stackrel{\mathrm{even}}{\longrightarrow}}(\C^p)^{\otimes k} = W$
and constructing the local evolution by gates acting on the target
space.  The imbeddings are named for the fact that in the usual
presentation of the braid group, the odd (even) numbered
generators can be implemented by restricting an action on $W$ to
image odd $V(\Sigma) \bigl(\text{even} V(\Sigma)
\bigr)$.
\\
\\
\noindent{\bf Proof:} The case genus $(\Sigma)=0$ with all boundary
components carrying identical labels (this contains the classical,
uncolored Jones polynomial case [J] [Wi]) is treated first. For
any number $q$ of punctures ($q = 10$ in the illustration) there
are two systematic ways of dividing $\Sigma$ into pants
($3-$punctured spheres) along curves
$\overset{\RI}{\alpha}=\{\alpha_1 , \dots, \alpha_{q-3}\}$ or
along $\overset{\RI}{\beta} = \{\beta_1,\dots,\beta_{q-3}\}$ so
that a sequence of $q$ $F$ moves ($6j-$moves in physics notation)
transforms $\overset{\RI}\alpha$ to $\overset{\RI}\beta$.

\vskip.2in \epsfxsize=3in \centerline{\epsfbox{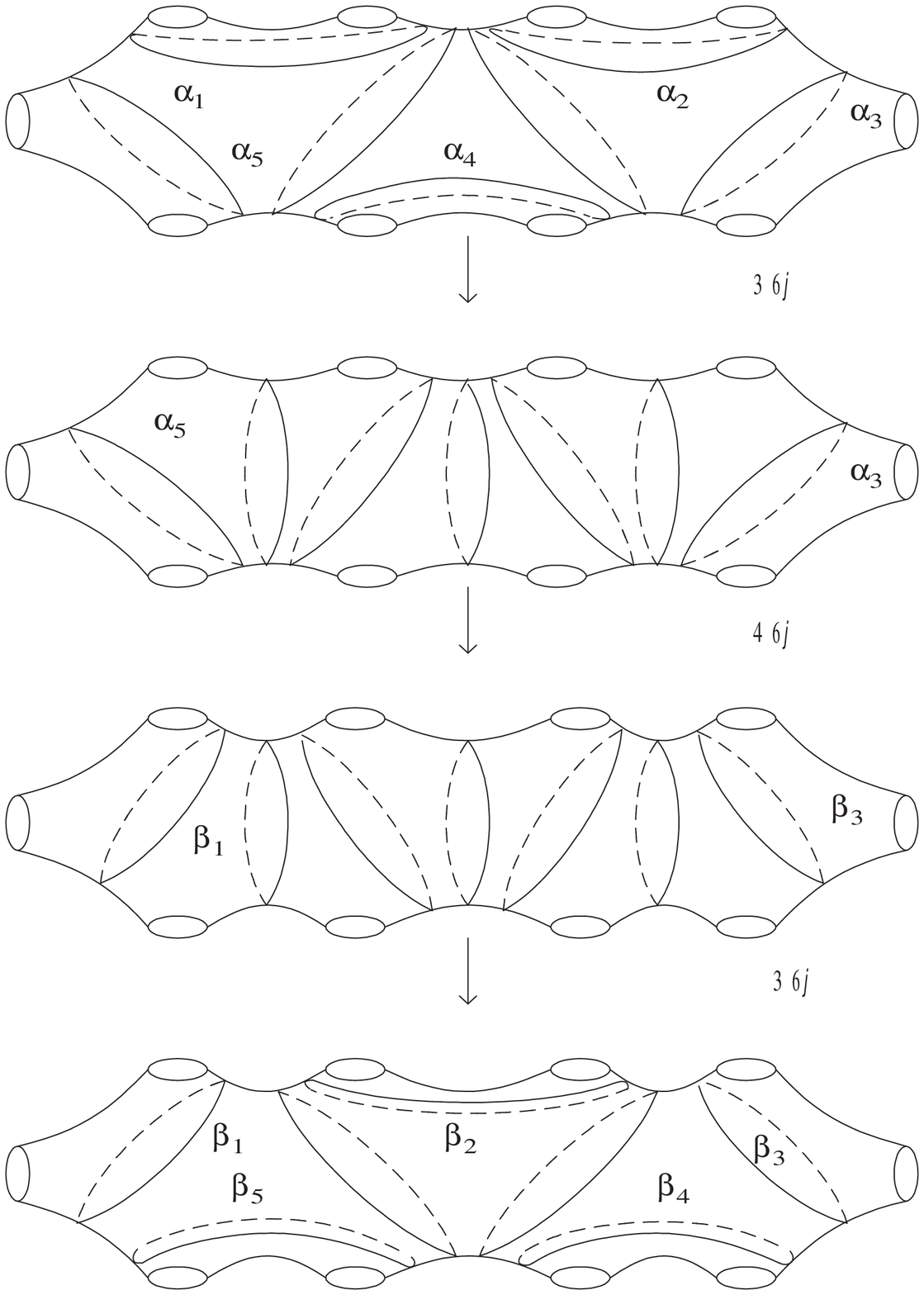}}
\centerline{Figure 3.} \vskip.2in

Let $X=\bigoplus_{(a, b, c) \epsilon \mathcal{L}^3} V_{abc}$ be
the orthogonal \underline{sum} of all \underline{sectors} of the
pants Hilbert space.  Distributing $\bigotimes$ over $\bigoplus$,
the tensor power $X^{\otimes (q-2)} := W$ is the sum over all
labelings of the Hilbert space for $\coprod (q-2)$ pants. Choosing
parameterizations, $W$ is identified with both the label sum space
$(\Sigma_{\text{cut}\overset{\RI}{\alpha}})$ and sum
$(\Sigma_{\text{cut}\overset{\RI}{\beta}}).$  Now $\Sigma$ is
assembled from the disjoint union by gluing along
$\overset{\RI}{\alpha}$ or $\overset{\RI}{\beta}$ so the gluing
axiom defines imbeddings $i(\overset{\RI}{\alpha})$ and
$i(\overset{\RI}{\beta})$ of $V(\Sigma, \ell)$ as a direct summand
of $X^{\otimes (q-2)}=W$.

Consider the action of braid move about $\alpha$.  This acts
algebraically as $\theta(\alpha_i)$ on a single $X$ factor of $W$
and as the identity on other factors.  This action leaves
$i(\overset{\RI}{\alpha})\,\,\bigl(V(\Sigma, \ell)\bigr)$
invariant and can be thought of as a $\lq\lq$qupit" gate:
\[\theta (\alpha_i) = V(\text{braid}_{\alpha_i}): X\rightarrow X\]
where dimension $dim(X)=p.$  Similarly the action of
$V(\text{braid}_{\beta_i})$ leaves $i(\overset{\RI}{\beta})$
invariant. It is well known [B] that the union of loops
$\overset{\RI}{\alpha} \cup \overset{\RI}{\beta}$ determines a
complete set of generators of the braid group.  The general
element $\omega$, which we must simulate by an action on $W$ is a
word in braid moves on $\alpha$'s and $\beta$'s. Part of the
\underline{basic data} - implied by the gluing axiom for a UTMF is
a fixed identification between elementary gluings:

\[ {F_{abcd}:\bigoplus_{x\epsilon\mathcal{L}} V_{xab}\otimes
V_{\widehat{x}c d} \longrightarrow
\bigoplus_{y\epsilon\mathcal{L}} V_{ybc} \otimes V_{\widehat{y}d
a}}
\]
corresponding to the following two decompositions of the
$4-$punctured sphere into two pairs of pants (The dotted lines are
pant $\lq\lq$seams", the uncircled number indicate boundary
components, the letters label boundary components, and the circled
numbers order the pairs of pants.):

\vskip.2in \epsfxsize=5in \centerline{\epsfbox{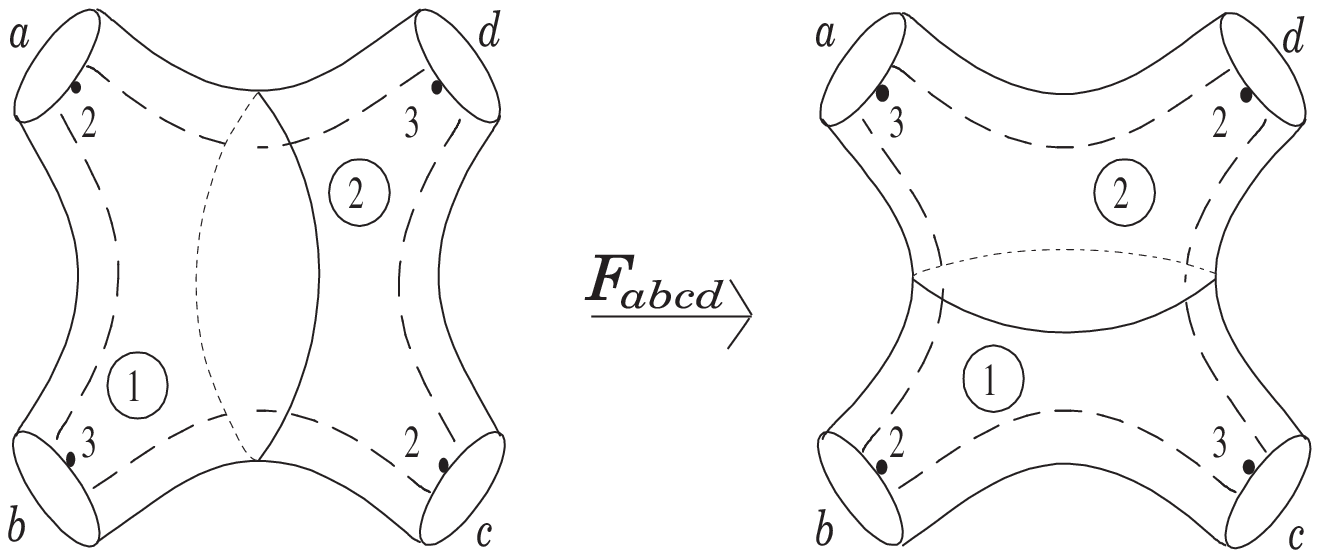}}
\centerline{Figure 4.} \vskip.2in

For each $F$, we choose an extension to a unitary  map
$F^{\prime}: X \bigotimes X \longrightarrow X \bigotimes X$. Then
extend $F^{\prime}$ to $ \overline{F}$ by tensoring with identity
on the $q-4$ factors unaffected by $F$. The composition of $q \,\,
F$'s, extended to $q\,\,\overline{F}$'s, corresponding to the $q$
moves illustrated in the case $q=10$ by Figure 3. (For $q>10$
imagine the drawings in figure 3 extended periodically.) These
define a unitary transformation $T:W \rightarrow W$ with $T \circ
i(\overset{\RI}{\alpha}) = i(\overset{\RI}{\beta})$. The word
$\omega$ in the braid group can be simulated by $\tau$ on $W$,
where $\tau$ is written as a composition of the unitary maps $T$,
$T^{-1}$, $ \theta(\alpha_{i})$, and $\theta(\beta_j)$. For
example,
\[\beta_5 \alpha_1 \beta_2^{-1}\alpha_1 \alpha_3\]
would be simulated as
\[\tau = T^{-1} \circ
\theta (\beta_5) \circ T \circ \theta (\alpha_1) \circ T^{-1}
\circ \theta(\beta_2^{-1}) \circ T \circ \theta(\alpha_1) \circ
\theta(\alpha_3).\]

As described $\tau$ has length $\leq 2q$ length $\omega$. The
dependence on $q$ can be removed by dividing $\Sigma$ into
$\frac{q}{2}$ overlapping pieces $\Sigma_i$, each $\Sigma_i$ a
union of 6 consecutive pants. Every loop of $\overset{\RI}{\alpha}
\cup \overset{\RI}{\beta}$ is contained well within some piece
$\Sigma_i$ so instead of moving between two fixed subspaces
$i_\alpha (V)$ and $i_\beta (V)\subset W$, when we encounter a
$\beta_j$, do constantly many $\overline{F}$ operations to find a
new pants decomposition modified locally to contain $\beta_j$.
Then $\theta(\beta_j)$ may be applied and the $\overline{F}$
operations reversed to return to the $\alpha$ pants decomposition.
The resulting simulation can be made to satisfy length $\tau \leq
7$ length $\omega$.  This completes the braid case with all
bounding labels equal - an important case corresponding to the
classical Jones polynomial [J].
\\

\noindent{\bf Proof of Lemma:} We have described the $F-$move on the
$4-$punc- \break tured sphere both geometrically and under the functor. The
$S-$move is between two pants decompositions on the punctured
torus $T^{-}$ (Filling in the puncture, a variant of $S$ may act
between two distinct annular decomposition of $T^2$.  We suppress
this case since, without topological parameter, there can be no
computational complexity discussion over a single surface).

\vskip.2in \epsfxsize=4in \centerline{\epsfbox{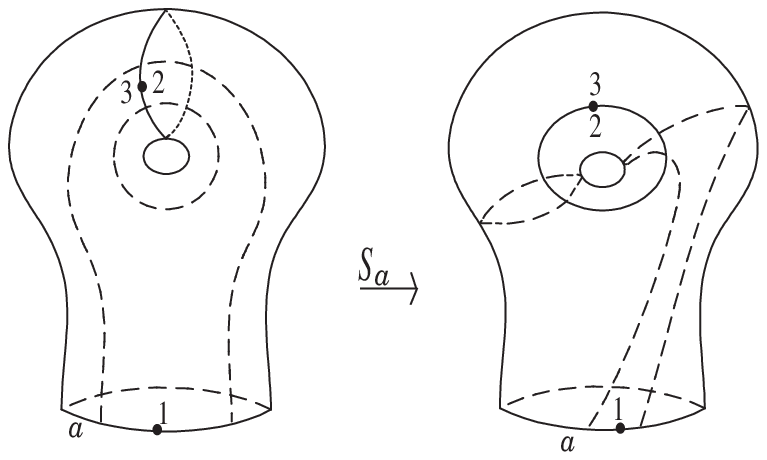}}
\centerline{\tn{Figure 5.} $ V(S):\underset{x \in
\mathcal{L}}{\bigoplus}  V_{a x \hat{x}}\longrightarrow
\underset{y \in \mathcal{L}}{\bigoplus} V_{ay\hat{y}}$ }
\vskip.2in

By [Li] or [HT] that one may move between any two pants
decompositions via a finite sequence of moves of three types: $F$,
$S$, and diffeomorphism $M$ supported on the interior of a single
pair of pants (appendix [HT]). To pass from $\mathcal{D}$, our
$\lq\lq$base point" decomposition,  to $\mathcal{D}_\omega $, $F$
and $S$ moves alone suffice and the logarithmic count is a
consequence of the $\log$ depth nest of cuff loops of
$\mathcal{D}$ on the planar surface obtained by cutting $\Sigma$
along type (1) $\gamma$ curves. Below we draw examples of short
paths of $F$ and $ S$ moves taking $\mathcal{D}$ to a particular
$\mathcal{D}_\omega$.
\\
The logarithmic count is based on the proposition.
\\
\begin{prop}
Let $K$ be a trivalent tree of diameter $=d$ and $f$ be a move,
which locally replaces \includegraphics[width=1cm,
height=.65cm]{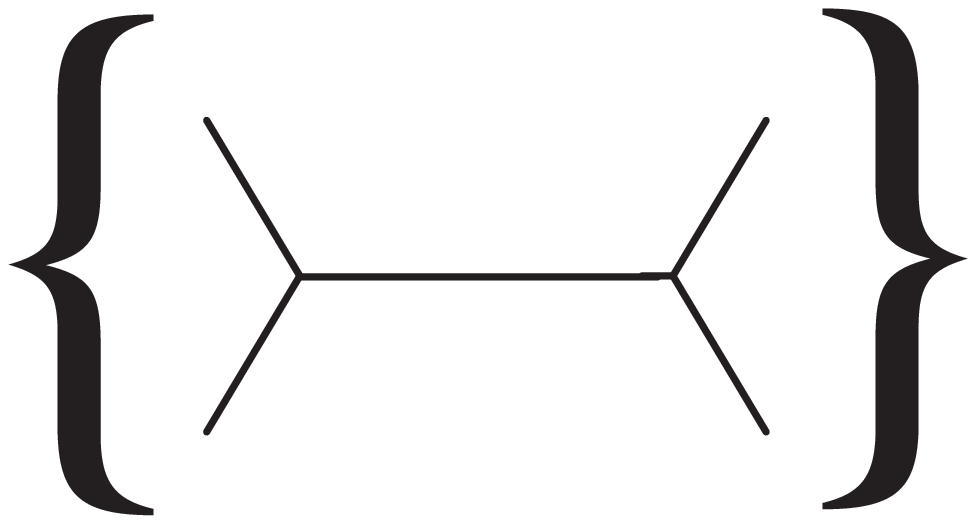} and with \includegraphics[width=1cm,
height=.65cm]{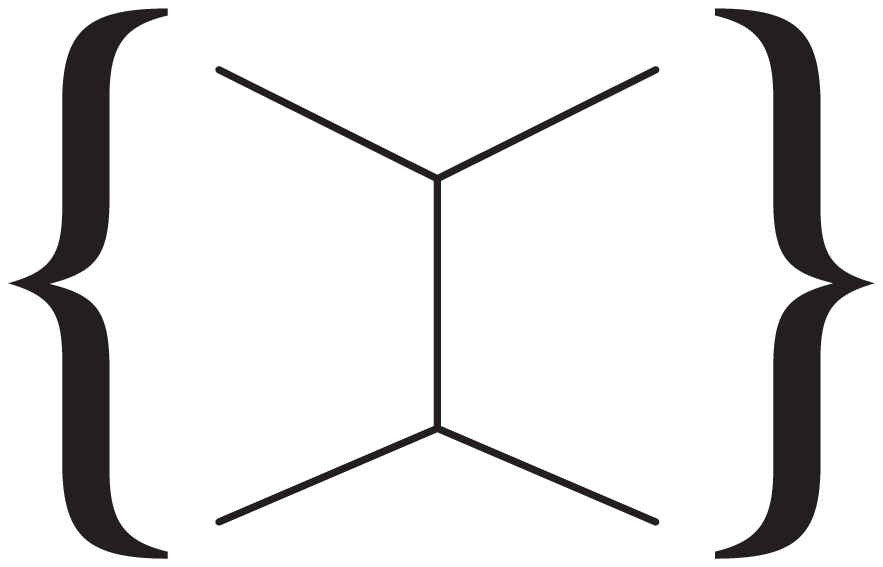}, then any two leaves of $K$ can be
made adjacent by $\leq d$ moves of type $f$.
\end{prop}

Passing from $K$ to a punctured sphere obtained by imbedding ($K$,
univalent vertices) into $(\frac{1}{2} R^{3}, R^{2}),$ thickening
and deleting the boundary $R^{2}$, the $f$ move induces the
previously defined $F$ move.$\square \square$
\\

Some example of paths of $F$, $S$ moves:

\vskip.2in \epsfxsize=6in \centerline{\epsfbox{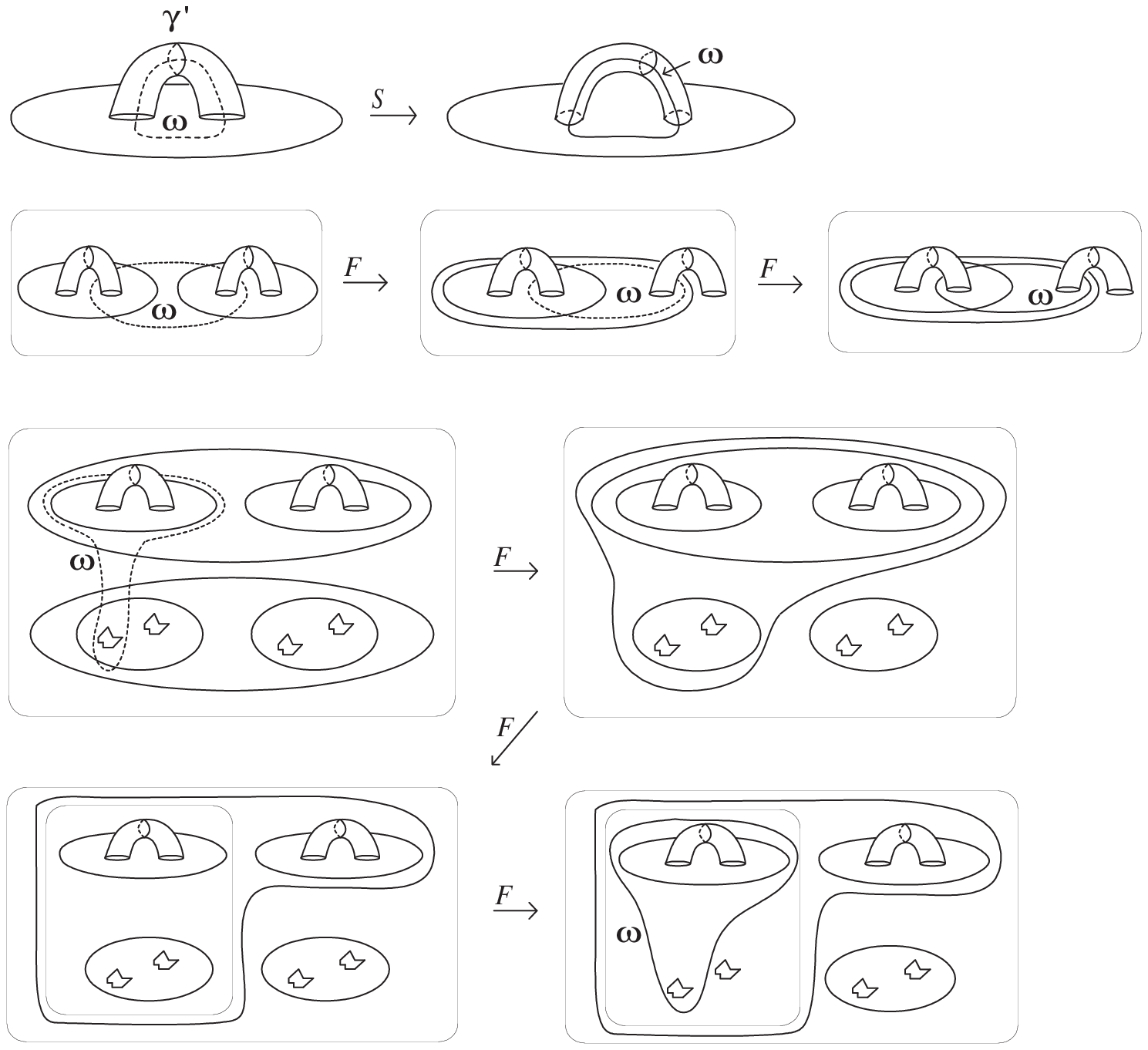}}
\centerline{Figure 6.} \vskip.2in

\noindent{\bf Continuation of the proof of the theorem}: For the general
case, we compute on numerous imbeddings of $V(\Sigma)$ into $W$
(rather than on two: $i_\alpha \big(V(\Sigma)\big)$ and $i_\beta
\big(V(\Sigma)\big)$ as in the braid case).  Each imbedding is
determined by a pants decomposition and the imbedding changes (in
principle) via the lemma every time we come to a new literal of
the word $\omega$.  Recall that $\omega\in\mathcal{M}$, the
mapping class group, is now written as a word in the letters (and
their inverses) of type $D_\gamma$, $D_\gamma'$, $D_\epsilon$, and
$B_\delta$.  Pick as a home base a fixed pants decomposition
$\mathcal{D}_0$ corresponding to $i_0 \big(V(\Sigma)\big)\subset
W$.  If the first literal is a twist or braid along the s.c.c.
$\omega$, then apply the lemma to pass through a sequence of $F$
and $S$ moves from $\mathcal{D}_0$ to $\mathcal{D}_{1}$ containing
$\omega$ as a $\lq\lq$cuff" curve. As in the braid case, choose
extensions $\overline{F}$ and $\overline{S}$ to unitary
automorphisms of $W$ and applying $V$ to the composition gives a
transformation $T_1$ of $W$ such that $i_1 = T_1 \circ i_0$, $i_1$
being the inclusion $V(\Sigma)\rightarrow W$ associated with
$\mathcal{D}_1$. Now execute the first literal $\omega_1$ of
$\omega$ as a transformation $\theta(\omega_1)$, which leaves
$i_1\big(V(\Sigma)\big)$ invariant and satisfies: $\theta
(\omega_i) \circ i_1 = i_1 \circ V(\omega_1)$.  Finally apply
$T_1^{-1}$ to return to the base inclusion
$i_0\big(V(\Sigma)\big)$. The previous three steps can now be
repeated for the second literal of $\omega$: follow $T_1^{-1}
\circ \theta(\omega_1) \circ T_1$ by $T_2^{-1} \circ
\theta(\omega_2) \circ T_2$. Continuing in this way, we construct
a composition $\tau$ which simulates $\omega$ on $W$:
\[ \tau = T_n^{-1} \circ \theta(\omega_n) \circ T_n^{-1} {\ddots} \circ
T_1^{-1} \circ \theta (\omega_1) \circ T_1 .
\]
From lemma 2.1 the length of this simulation by one
$\big($corresponding to $S$ and $\theta(\omega_i)\big)$ and two
(corresponding to $F$ moves) qupit gates is proportional to
$n=$length $\omega$ and $ \log b_1 (\Sigma)$, where $p =\dim(X)$.
\\
\\
\noindent{\bf Proof of Extension:} What is at issue is the
number of preparatory moves to change the base point decomposition
$\mathcal{D}$ to $\mathcal{D}_{\gamma}$ containing $ \gamma =
\alpha_i$ or $\beta_i$ as a cuff curve $1 \leq i \leq k$ or $j$.
We have defined the $F$ and $S$ moves rigidly, i.e. with specified
action on the seams.  This was necessary to induce a well defined
action on the functor $V$.  Because of this rigid choice, we must
add one more move - an $M$ move - to have a complete set of moves
capable of moving between any two pants decompositions of a
surface (compare [HT]).  The $M$ move is simply a Dehn twist
supported in a pair of pants of the current pants decomposition;
it moves the seams (compare chapter 5 [Wa]). Note that if $M$ is a
$+1$ Dehn twisit in a s.c.c. $\omega$ then, under the functor,
$V(M)$ is a restriction of $\theta(\omega)$ in the notation
above.

As in [HT], the cuff curves of $\mathcal{D}$ may be regarded as
level curves of a Morse function $f: \Sigma \rightarrow R^{+}$,
constant on boundary components which we assume to have minimum
complexity (= total number of critial points) satisfying this
constraint. Isotope $\alpha$ (we drop the index) on $\Sigma$ to
have the smallest number of local maximums with respect to $f$ and
is disjoint from critical points of $f$ on $\Sigma$.

Now generically deform $f$ in a thin annular neighborhood of
$\gamma$ so that $\gamma$ becomes a level curve.  Consider the
graphic $G$ of the deformation $f_t$, $0 \leq t \leq 1$.  For
regular $t$ the Morse function $f_t$ determines a pants
decomposition:  let the $1$- complex $K$ consist of $\Sigma/\sim$
where $x \sim y$ if $x$ and $y$ belong to the same component of a
level set of $f_t$,  and let $L \subset K$ be the smallest complex
to which $K$ collapses relative to endpoints associated to
boundary components. For example in figure 8, the top tree does
not collapse at all while in the lower two trees the edge whose
end is labeled, $\lq\lq$local max" is collapsed away.  The
preimage of one point from each intrinsic $1-$cell of $L$ not
containing a boundary point constitutes a $\{\text{cuffs}\}$
determining a pants decomposition $\mathcal{D}_t$. For singular
$t_0$, let $\mathcal{D}_{t_{0} - \epsilon}$ and
$\mathcal{D}_{t_{0} + \epsilon}$ may differ or may agree up to
isotopy.  The only change in $\mathcal{D}$ occurs when $t$ is a
crossing point for index$=1$ handles where the two critical points
are on the same connected component of a level set
$f_{t}^{-1}(r)$.  There are essentially only three possible
$\lq\lq$Cerf-transitions" and they are expressible as a product of
1, 2, or 3 $F$ and $S$ moves together with braid moves whose
number we will later bound from above.  The Cerf transitions on
$\mathcal{D}$ are shown in Figure 7, together with their
representation as compositions of elementary moves.

\vskip.2in \epsfxsize=6in \centerline{\epsfbox{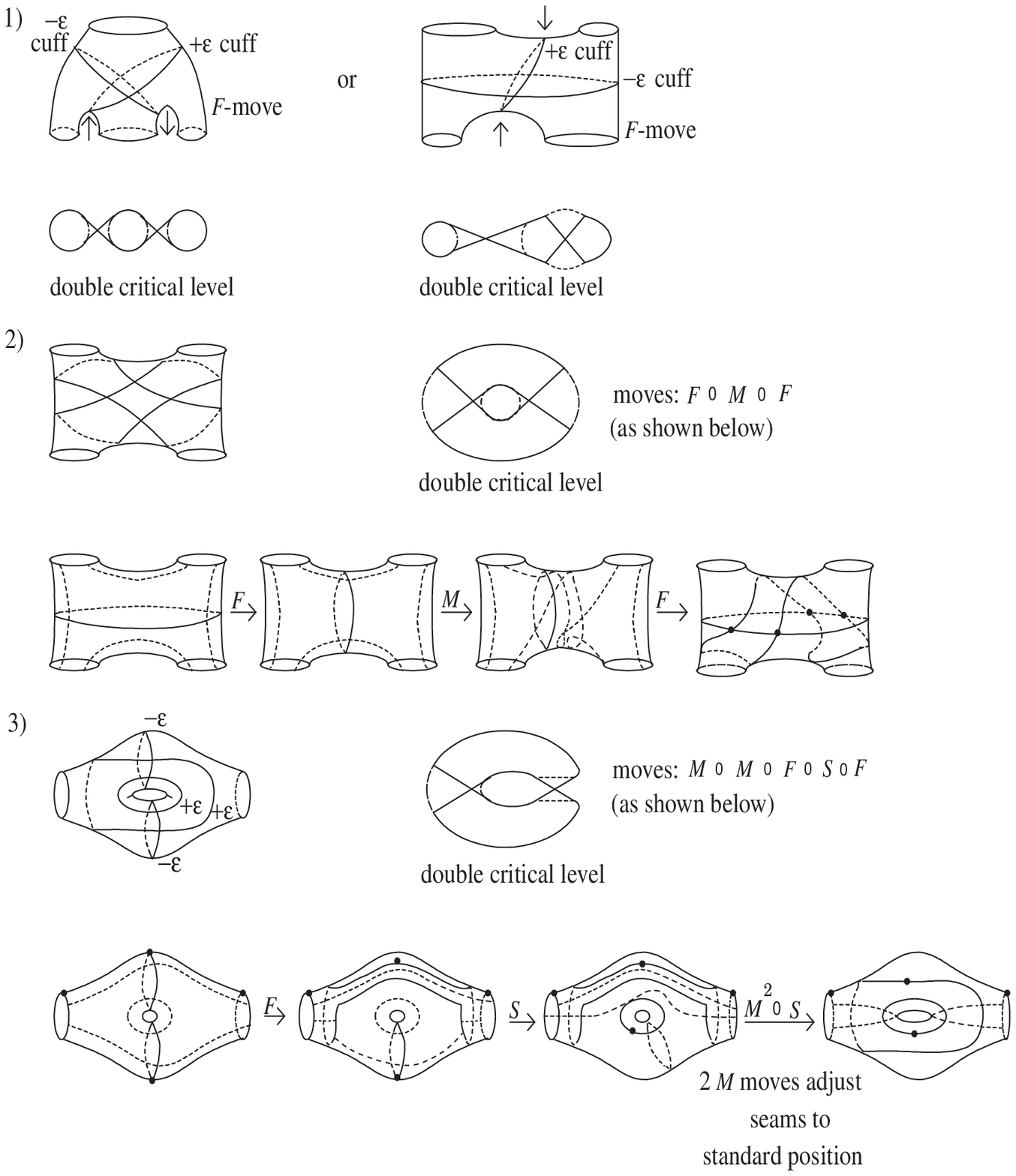}}
\centerline{\tn{Figure 7.}} \vskip.2in

Critical points of $f|_{\gamma}$ become critical points of $f_t$
of the same index once the deformation as passed an initial
$\epsilon_0 > 0$, and before any saddle-crossings have occurred.
Let $P$ be a pant from the composition induced by $f$ and $\delta
\subset \gamma \cap P$ an arc.  Applying the connectivity
criterion of the previous paragraph, we can see that flattening a
local maxima can effect at most the two cuff circles which
$\delta$ meets, and these by elementary Cerf transition shown in
Figure 8.

\vskip.2in \epsfxsize=4in \centerline{\epsfbox{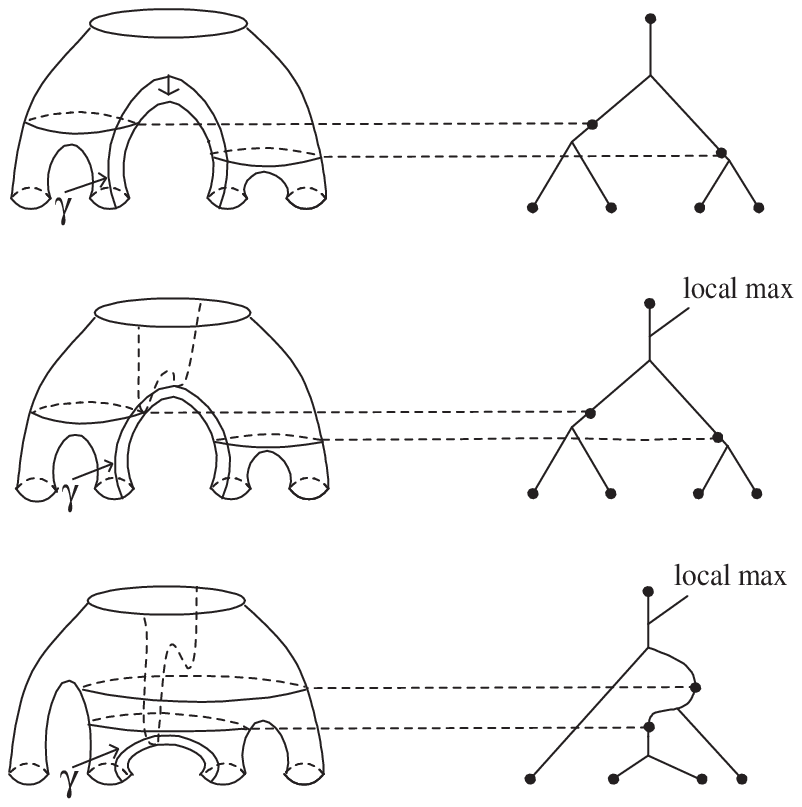}}
\centerline{Figure 8. Pulling $\gamma$ down yields $F \circ
F$}
\vskip.2in

If $\gamma$ crosses the seam arcs then the transitions are of the
Cerf type, precomposed with $M-$moves to remove these crossings as
shown in Figure 9. Dynamically seam crossings by $\gamma$ produce
\underline{saddle} \underline{connections} in the Cerf diagram.

\vskip.2in \epsfxsize=4in \centerline{\epsfbox{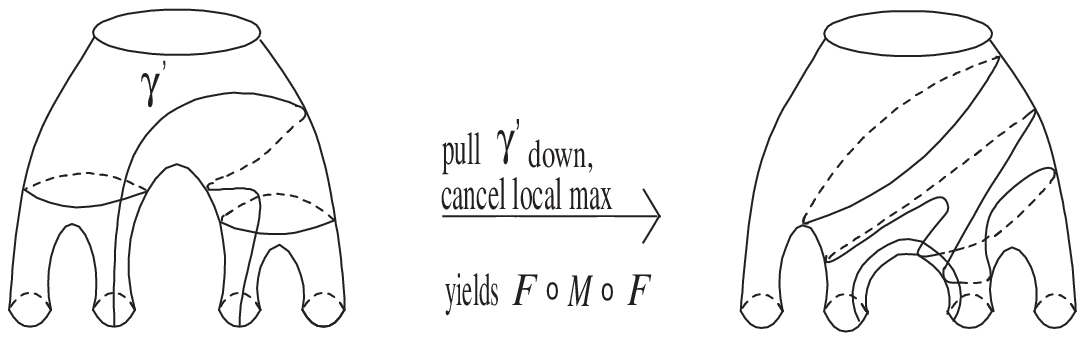}}
\centerline{\tn{Figure 9.}} \vskip.2in

The total number of these twists is bounded by length $(\gamma)$.
The number of flattening moves as above is less than or equal
$|\gamma \cap \text{cuffs}| \leq \text{length}(\gamma)$.  The
factor of 11 in the statement allows up to 5 $F$, $S$, and $M$
moves for expressing each Cerf singularity which arises in passing
from $\mathcal{D}_\circ$ to $\mathcal{D}_{\gamma}$ and the same
factor of 5 to pass back from $\mathcal{D}_\gamma$ to
$\mathcal{D}_\circ$ again, while saving at least one step to
implement the twist or build move along $\gamma$. This completes
the proof of the extension. $\square$
\\
\\
We should emphasize that although, we have adopted an
$\lq\lq$exact" model for the operation of the UTMF, faithful
simulation as derived above does not depend on a perfectly
accurate quantum circuit.  Several authors have proved a threshold
theorem [Ki], [AB], and [KLZ]: If the rate of large errors acting
on computational qubits (or qupits) is small enough, the size of
ubiquitous error small enough, and both are uncorrelated, then
such a computational space may be made to simulate with
probability $\geq \frac{2}{3}$ an exact quantum circuit of length
$=L$. The simulating circuit must exceed the exact circuit in both
number of qubits and number of operations by a multiplicative
factor $\leq$ poly $(\log L)$.
\section{Simulating TQFT's}
We conclude with a discussion about the three dimensional
extension, the TQFT of a UTMF.  In all known examples of TMF's
there is an extension to a TQFT meaning that it is possible to
assign a linear map  $V(\Sigma)\overset{b_\ast}{\rightarrow}
V(\Sigma')$ subject to several axioms [Wa] and [T] whenever
$\Sigma$ and $\Sigma'$ cobounds a bordism $b$ (with some
additional structure).  The case of bordisms with a product
structures is essentially the TMF part of the theory. Unitarity is
extended to mean that if the orientation of the bordism $b$ is
reversed to $\overline{b}$, we have $b_\ast^\dagger =
(\overline{b})_\ast$. It is known that a TMF has at most one
extension to a TQFT and conjectured that this extension always
exists.  Non-product bordisms correspond to some loss of
information of the state. This can be understood by factoring the
bordism into pieces consisting of a product union a $2-$handle:
$\Sigma \times I \cup h$.  The $2-$handle $h$ has the form $(D^2
\times I,\,\, \partial D^2 \times I)$ and is attached along the
subspace $\partial D^2 \times I$. The effect of attaching the
handle will be to $\lq\lq$pinch" off an essential loop $\omega$ on
$\Sigma$ and so replace an annular neighborhood of $\omega$ by two
disks turning $\Sigma$ into a simpler surface $\Sigma'$.  It is an
elementary consequence of the axioms that if $b=\Sigma  \times I
\cup h$ then $b_\ast$ is a projector as follows:  Let
$\mathcal{D}$ be a pants decomposition containing $\omega$ as a
dissection curve. There are two cases:
\begin{enumerate}
\item $\omega$ appears as the first and second boundary
components of a single pant called $P_0$ or
\item$\omega$ appears as the first boundary component on two
distinct pants called $P_1$ and $P_2$.
\end{enumerate}
\[
\begin{array}{l}
V(\Sigma)\ =\smallskip\\ =\ \bigoplus\limits_{c \epsilon
\mathcal{L}} \Bigg( \biggl(
\bigoplus\limits_{a\epsilon\mathcal{L}}V_{a\hat{a}c} \biggr)
\bigotimes V \bigl( \Sigma\backslash P_{0}, \text{with label c
on}\;\partial_3 P_{0} \bigr) \Bigg),\ \text{case (1),}
\smallskip\\
\mathrm{or}
\smallskip\\
=\ \bigoplus\limits_{\mathrm{labels}} \biggl( \bigoplus\limits_{a
\epsilon \mathcal{L}} V_{abc} \bigotimes V_{\hat{a}d e}\biggr)
\bigotimes V\biggl(\Sigma\backslash \bigl(P_{1} \cup P_{2}\bigr),
\text{appropriate labels} \biggr),\ \text{case (2)}.
\end{array}
\]

In case (2), there may be relation $b=\hat{c}$ and/or $d=\hat{e}$
depending on the topology of $\mathcal{D}$.  The map $b_\ast$ is
obtained by extending linearly from the projections onto summands:

\begin{eqnarray*}
\bigoplus_{a, c\,\mathcal{L}}V_{a\hat{a}c} & \longrightarrow &
V_{1 1 1} \overset{\mathrm{canonically}}{\cong} V_{1} \quad \quad
\qquad \qquad \qquad \text{(case 1)} \\ \mathrm{or}
\\
\bigoplus_{a, b, c, d, e\, \epsilon \mathcal{L}}V_{abc} \bigotimes
V_{\hat{a}d e} & \longrightarrow & V_{1 b\hat{b}}\bigotimes V_{1
d\hat{d}} \overset{\mathrm{canonically}}{\cong}V_{b\hat{b}}
\bigotimes V_{d\hat{d}}\quad \text{(case 2)}
\end{eqnarray*}

If the orientation on $b$ is reversed the unitarity condition
implies that $\overline{b}$ determines an injection onto a summand
with a formula dual to the above.  Thus, any bordism's morphism
can be systematically calculated.

In quantum computation, as shown in [Ki], a projector corresponds
to an intermediate binary measurement within the quantum phase of
the computation, one outcome of which leads to cessation the other
continuation of the quantum circuits operation.  Call such a
probabilistically abortive computation a \underline{partial
computation} on a \underline{partial quantum circuit}.  Formally,
if we write the identity as a sum of two projectors: $\text{id}_V
= \Pi_0 + \Pi_1$, and let $\U_0$ and $\U_1$ be unitary operators
on an ancillary space $A$ with $\U_0(|0\rangle)= |0\rangle$ and
$\U_1 |0\rangle = |1\rangle$. The unitary operator $\Pi_0 \otimes
\U_0 + \Pi_1 \otimes \U_1$ on $V \otimes A $ when applied to
$|v\rangle \otimes |0\rangle$ is $| \Pi_0 v\rangle \otimes
|0\rangle + | \Pi_1 v\rangle \otimes |1\rangle$ so continuing the
computation only if the indicator $|0\rangle \in A$ is observed
simulates the projection $\Pi_0$.

It is clear that the proof of the theorem can be modified to
simulate $2$-handle attachments as well as Dehn twists and braid
moves along s.c.c.'s $\omega$ to yield:

\begin{SC}  Suppose $b$ is an oriented bordism from $\Sigma_0$ to
$\Sigma_1$, where $\Sigma_i$ is endowed with a pants decomposition
$\mathcal{D}_i$.  Let complexity $(b)$ be the total number of
moves of four types:  $F$, $S$, $M$, and attachment of a
$2$-handle to a dissection curve of a current pants decomposition
that are necessary to reconstruct $b$ from $(\Sigma_0 ,
\mathcal{D}_0 )$ to $(\Sigma_1 , \mathcal{D}_1)$. Then there is a
constant $c'(V)$ depending on the choice of UTQFT and $p(V)$ as
before (for the TQFTs underlining TMF) so that $b_\ast :V(\Sigma_0
)\rightarrow V (\Sigma_1 )$ is simulated (exactly) by a partial
quantum circuit over ${\C}^p$ of length $\leq c'$ complexity
$(b)$.
\end{SC}

\end{document}